\documentclass[10pt]{article}
\usepackage[english]{babel}
\usepackage{mathpazo}
\usepackage{graphicx}
\usepackage{amsmath,amssymb,bm}
\usepackage{cite}
\usepackage[ansinew]{inputenc}
\usepackage{xcolor}
\definecolor{darkblue}{rgb}{0,0,0.7}
\usepackage[breaklinks=true,colorlinks=true,linkcolor=darkblue,citecolor=darkblue,urlcolor=darkblue]{hyperref}
\usepackage{geometry}
\usepackage{breakurl}
\usepackage{caption}
\usepackage[symbol*]{footmisc}
\usepackage{etoolbox}

\DefineFNsymbols*{asdf}{*\#*}
\setfnsymbol{asdf}

\appto\UrlBreaks{\do\-}

\title{(In)Feasability of Studying Ultra-High-Energy\\ Cosmic Rays
  with Smartphones}
\author{Michael~Unger$^{1,2,}$\footnote{\href{mailto:mu495@nyu.edu}{mu495@nyu.edu}}$\;\;$
  and Glennys
  Farrar$^{1,}$\footnote{\href{mailto:gf25@nyu.edu}{gf25@nyu.edu}}\\\small
  $^{1}$Center for Cosmology and Particle Physics, New York University
  \\\small $^{2}$Institut f\"ur Kernphysik, Karlsruhe Institute of
  Technology} \date{May 18, 2015}

\begin{document}

\maketitle

\begin{abstract}
We estimate the effective area available for cosmic-ray detection with
a network of smartphones under optimistic conditions.  To measure
cosmic-ray air showers with a minimally-adequate precision and a detection
area similar to existing ground-based detectors, the fraction of
participating users needs to unrealistically large.  We conclude that
the prospects of cosmic-ray research using smartphones are very
limited.
\end{abstract}

\vspace*{0.5cm} Interactions of ultra-high-energy cosmic rays (UHECRs)
with nuclei of the Earth's atmosphere create an avalanche of secondary
charged particles that can be detected on the surface by giant air
shower arrays such as the Pierre Auger Observatory~\cite{Aab:2015zoa}
or the Telescope Array~\cite{AbuZayyad:2012kk}.  Recently it has been
proposed to use the camera sensors of smartphones to detect these air
showers. The Distributed Electronic Cosmic-Ray Observatory ({\scshape
  Deco})~\cite{deco} aims to use {this technique} for educational
purpose, whereas the {\scshape Crayfis} group (Cosmic Rays Found In
Smartphones) proposes to { employ smartphones} for scientific
research~\cite{crayfis, Whiteson:2014kca}.

Since the flux of UHECRs is very small (of the order of 1 per km$^2$
per century), an essential feature for a cosmic-ray observatory is to
cover a sufficiently large detection area.  In the present short note
we study this most basic feasibility requirement for studying
ultra-high-energy cosmic rays with an array of smartphones.
Additional important issues, if the achievable event rate is
sufficient for such a detector to be of scientific interest, are the
capability to trigger on air showers, the quality of the event
reconstruction, and the reconstruction uncertainties in energy and
arrival direction (see also~\cite{physworld}).

The effective area of an array of smartphones {\itshape \`a la} {\scshape
  Crayfis}~\cite{Whiteson:2014kca} is given by the total area on the
globe where the density of phones participating in the UHECR search
exceeds the threshold for cosmic ray detection, weighted by the duty
cycle $f_{\rm up}$ of each phone,
\begin{equation}
  A_{\rm eff} = f_{\rm up}\, \sum_i A_i
\, {\mathcal H}(f_{{\rm sp}} \, f_{{\rm app}}\,\rho_i - \rho_{\rm thr}).
\label{eq:area}
\end{equation}
Here the sum runs over all grid points on the surface of Earth,
${\mathcal H}$ is the Heaviside function, $\rho_i$ denotes the
population density in bin $i$ and the area of bin $i$ is given by
$A_i$.  $f_{\rm sp}$ and $f_{\rm app}$ are the fraction of the
population owning smartphones and running the app respectively.  For
simplicity, $f_{\rm sp}$ and $f_{\rm app}$ are assumed to be
{globally constant.}

\begin{figure}[t!]
\centering
 \includegraphics[width=0.74\linewidth]{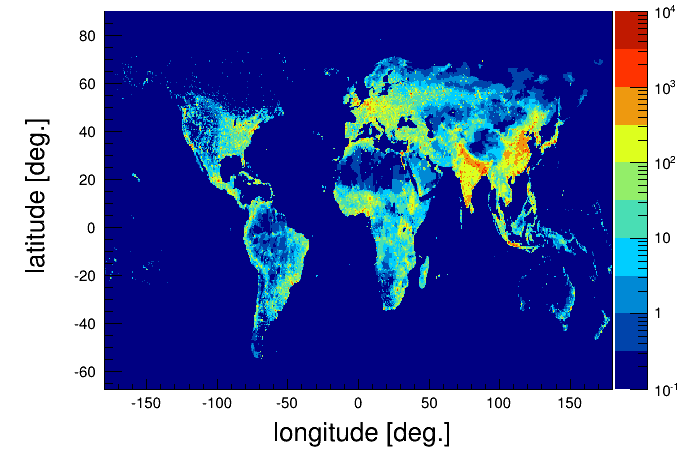}
 \caption{\texttt{GRUMPv1} population density averaged over
   $0.5^\circ\times0.5^\circ$ bins. Densities are capped at ${\rm
     max}({\rm min}(\rho, 10^{-1}), 10^{4})$.}
 \label{fig:grid}
\end{figure}

Estimates for the population density $\rho_i$ are provided by the
Global Rural-Urban Mapping Project (\texttt{GRUMPv1}) for the year
2000 on a 30 arcsecond grid with an average area per grid point of
0.6~km$^{2}$~\cite{grump}. The corresponding population density map is
shown in Fig.~\ref{fig:grid}.

The minimal phone density, $\rho_{\rm thr}$, for detecting UHECR
showers with an acceptable energy resolution of $\Delta E / E \lesssim
30\%$ is 5000 phones per km$^2$, for a threshold energy of $E >
10^{19}$~eV, and 1000 phones per km$^2$ for $E > 10^{20}$~eV, assuming
an optimistic value for the effective sensor area of $A\,\varepsilon = 5
\times 10^{-5}$~m$^2$ (cf.\ Fig.~7 in~\cite{Whiteson:2014kca}).~\footnote{It
should be noted that even in this case, the energy and angular
resolution that could be achieved by {\scshape Crayfis} is much worse
than for conventional air shower arrays. For instance, the energy
resolution of the Pierre Auger Observatory is 12\% and
the angular resolution is better than $1^\circ$~\cite{Aab:2015zoa}. The arrival
direction resolution estimated by the {\scshape Crayfis} group would be about 20$^\circ$
(10$^\circ$) in azimuthal (zenith) angle for events with
$E=10^{20}$~eV, and much worse for those at $10^{19}$
eV~\cite{Whiteson:2014kca}.}

The number of active smartphones depends on the fraction of phones
running at any given time, $f_{\rm up}$.  Assuming that {\scshape
  Crayfis} is running overnight while the phone is charging, $f_{\rm
  up} = \frac{1}{3}$ seems to be a plausible value. Since for each
grid point the down time is correlated between phones (no data taking
during day time), $f_{\rm up}$ diminishes the overall duty cycle, but
does not enter into the Heaviside function in Eq.~(\ref{eq:area}).

The fraction of people owning a smartphone, $f_{\rm sp}$, varies
considerably from country to country. Values for 2013 are 56\% for the
USA, 47\% for China, 17\% for India and 14\% for
Indonesia~\cite{mobile}.  In the following we adopt 50\% as an
optimistic estimate that might be reached in the future.

The most uncertain ingredient of the calculation is the fraction of
smartphone owners that install {\scshape Crayfis}, $f_{\rm app}$.  As
a point of reference, the three popular astronomical apps {\it SkEye,
  Star Tracker} and {\it Google Sky Map} accumulate 11-56 million
installations together~\cite{star}. However, Citizen Science apps,
such as {\it CrowdMag}~\cite{crowdmap} by NOAA for the monitoring of
the Earth magnetic field, can have as few as 1000 - 5000
installations.  And even the highly popular desktop-based search for
extraterrestial life, SETI@home, has only a little more than 100,000
active users~\cite{seti}.

\begin{figure}[t!]
\centering
 \includegraphics[width=0.72\linewidth]{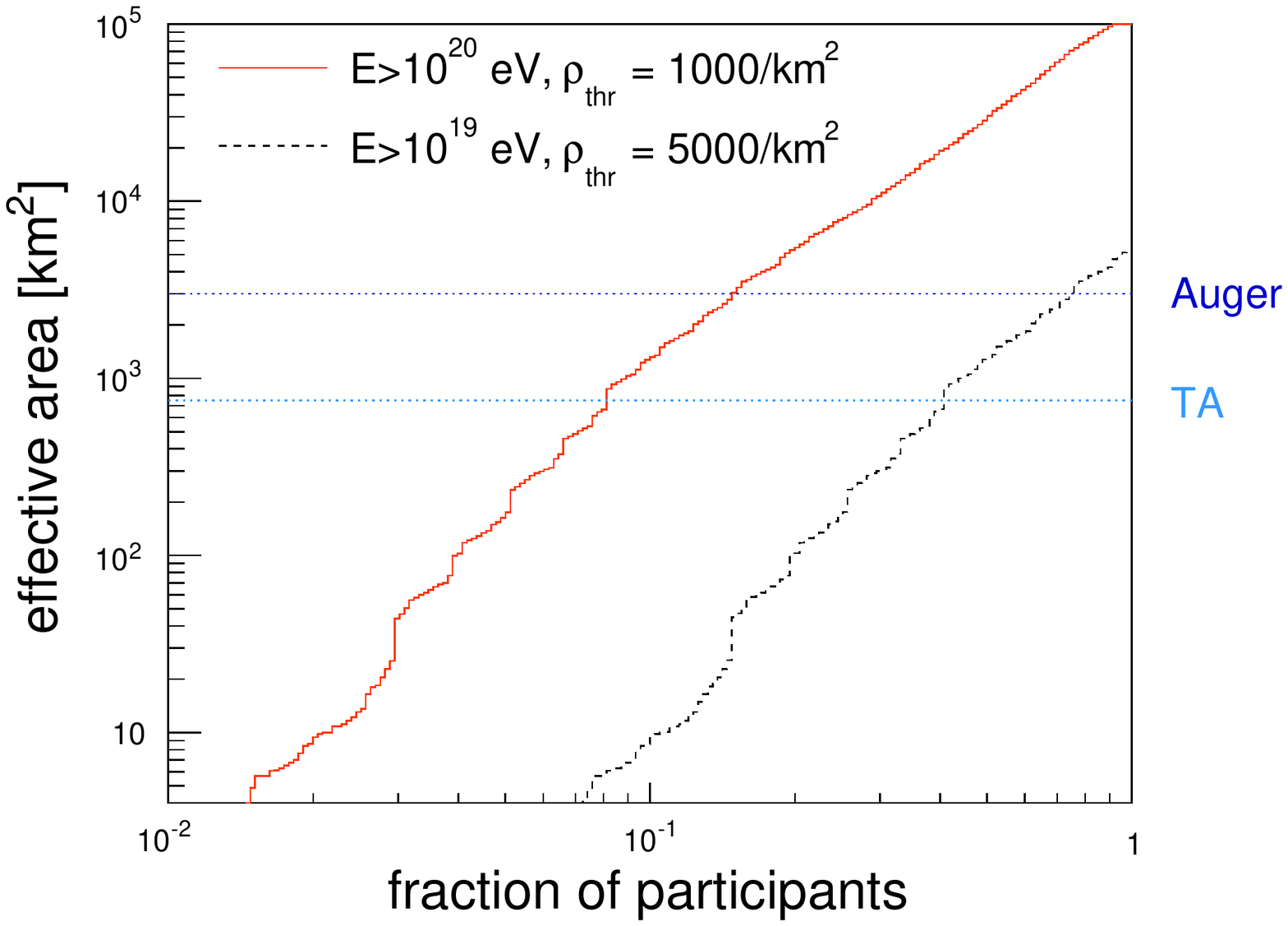}
 \caption{Effective area of {\scshape Crayfis} for two energy
   thresholds as a function of the fraction of smartphone users
   participating in the UHECR search. Horizontal lines indicate the
   area of the surface detectors of the Pierre Auger Observatory and
   the Telescope Array.}
 \label{fig:area}
\end{figure}

The effective area of {\scshape Crayfis} as a function of $f_{\rm
  app}$ is shown in Fig.~\ref{fig:area} for two values of $\rho_{\rm
  thr}$ corresponding to phone densities that allow the detection of
air showers above $10^{19}$~eV and $10^{20}$~eV.  As can be seen, to
reach the effective area of of the Pierre Auger Observatory, {\scshape
  Crayfis} would have to run with a participation rate of $f_{\rm app}
> 15\%$ if it targeted only very energetic showers above
$10^{20}$~eV. For a lower energy threshold of $10^{19}$~eV, the
participation rate would have to be $f_{\rm app} > 75\%$

For comparison, an upper limit of $f_{\rm app}$ for the aforementioned
popular astronomical apps is $0.0056$ assuming that 1~billion devices
running Android OS~\cite{android} and that the installation count
reflects the actual number of apps in use.  Even if the hunt for
cosmic rays above $10^{20}$~eV were as popular as stargazing, the
corresponding effective area is zero, because there is no spot on
Earth where the population density exceeds $\rho_{\rm thr} / (f_{\rm
  sp} \, f_{\rm app}) = 4\times 10^{5}$ people per 1~km$^{2}$.\\

We conclude therefore that unless the popularity and usage of the
{\scshape Crayfis} app significantly exceeds that achieved by current
popular science apps, there is little hope of { contributing to the
  science of ultra-high energy cosmic rays} with the smartphone
approach.\\

\newpage
\noindent
{\bf Acknowledgments:}
This work is supported by the EU-funded Marie Curie Outgoing
Fellowship, Grant PIOF-GA-2013-624803, U.S. National Science
Foundation grant NSF-PHY-1212538, and the James Simons Foundation.

\sloppy

\end{document}